# Superparamagnetic iron oxide nanoparticles conjugated with doxorubicin for targeting breast cancer

**Authors:**

Enrico Catalano

Jürgen Geisler

Vessela N. Kristensen

Department of Clinical Molecular Biology (EpiGen), Akershus University Hospital, University of Oslo (UiO), Oslo, Norway. email: enrico.catalano@medisin.uio.no

**Corresponding author**

Catalano Enrico

Department of Clinical Molecular Biology (EpiGen), Akershus University Hospital, University of Oslo (UiO), Akershus University Hospital Sykehusveien 25 Epigen 1478 Lørenskog, Norway.

## Abstract

Development of the next generations of cancer therapy modalities is currently a crucial requirement of oncology. Advances in nanotechnology are bringing us closer to the development of dual and multifunctional nanoparticles that are challenging the traditional distinction between diagnostic and treatment agents. The chase of innovative, multifunctional, more efficient, and safer treatments is a major challenge in preclinical nanoparticle-mediated thermotherapeutic research for breast cancer. Here, we report that iron oxide nanoparticles have the dual capacity to act as both magnetic and drug delivery agents.

The aim of this work was to investigate the in vitro effect of the loading of doxorubicin (DOX) on negatively charged polycarboxylic iron-oxide nanoparticles (SPIONs) and Rhodamine B functionalized SPIONs on breast carcinoma cell lines. For proper analysis and understanding of cell behavior after administration of DOX-SPIONs compared with free DOX, a complex set of in vitro tests, including production of MTT assay, cell cycle determination, and cellular uptake, were utilized. In summary, we have developed a magnetic nanoparticle-based drug delivery system that sequentially delivers the cytotoxic drug doxorubicin to breast cancer cells (MCF-7 and MDA-MB-231). The drug-coated nanoparticles, DOX-NPs, were assembled stepwise, with doxorubicin adsorbed to bare iron oxide nanoparticles first, by electrostatic reaction and allowed for the complexation of doxorubicin. DOX-NPs were stable in solution at 37 °C and physiological pH (7.4). Rapid in vitro release of doxorubicin followed by gradual release of doxorubicin was triggered in aqueous solution by low pH (5.4) and heating. As a result of minimal internalization, the particles were not significantly toxic to noncancerous cells (MCF-10A). In contrast, they were internalized to a much greater extent in MCF-7 and MDA-MB-231 cells and were cytotoxic due to the synergistic action of the two drugs and the effects of hyperthermia. The drug-coated particles were able to inhibit growth and proliferation of breast cancer cells in vitro, indicating that the system has potential to act as an antimetastatic chemothermotherapeutic agent.

**Introduction**

Breast cancer is still one of the leading causes of cancer-related deaths in women in Europe and worldwide [1]. Due to the implementation of improved early diagnostic methods, modern surgery and more effective drugs in the adjuvant setting, the average 5-year survival rate has increased to about 90%. However, if distant metastasis occurs, patients are still evaluated as non-curable and have in general only a few years to live while on permanent drug therapy. Thus, current drug therapy is far from optimal, especially in the metastatic setting. Nanomedicine is a medicine made by nanotechnology. Drug delivery is one of the most important properties of nanomedicine. To increase the capacity of targeting delivery of anticancer drugs to tumors, nanoparticles are usually functionalized with targeted antibodies, peptides or other biomolecules.

Nanomedicine is a promising therapeutic strategy for breast cancer treatment. Nanomedicine products such as Doxil® and Abraxane® have already been extensively used for breast cancer adjuvant therapy with favorable clinical outcomes. However, these products were originally designed for generic anticancer purpose and not specifically for breast cancer treatment. With better nanoengineering and nanotechnology action against breast cancer, a number of novel promising nanotherapeutic strategies and devices can be implemented.

The growing interest in applying nanotechnology to cancer is largely attributable to its uniquely appealing features for drug delivery, diagnosis and imaging, synthetic vaccine development and miniature medical devices, as well as the therapeutic nature of some nanomaterials themselves [2]. The original intended use of nanomedicine is to improve human health. Nanosized magnetic nanoparticles are some of the best candidates for the development of more efficient methods for the synthesis of nanodrug delivery vehicles based on iron-oxide nanoparticles. Nanotheranostics, the integration of diagnostic and therapeutic function in one system using the benefits of nanotechnology, is extremely attractive for personalized medicine [2]. As the conventional chemotherapy is administered systemically, it often causes considerable adverse reactions such as nausea, hair loss, and bone marrow suppression, as well as liver and kidney toxicity [3]. These aspects determine the dose of the chemotherapeutic agents and limit their effects on the tumor site [3].

Many different novel cancer therapeutic strategies that have been explored in the last two decades, therapies that are responsive to external physical stimuli coupled with nanoparticles (NPs) that are capable of responding to light, magnetic fields, ultrasound, radio-frequency, or x-ray, have attracted substantial interest from many researchers across the world [17].

Magnetic fields (MF) are non-invasive in nature and have excellent human tissue penetration. Owing to this characteristic of MF, it has been used in clinics for whole body MRI. In recent years, MF has been used to activate drug delivery into the tumor region through field dependent thermal and non-thermal effects. High frequency (>10 kHz) alternating magnetic fields (AMF) responsive MNPs can generate heat by utilizing different physical mechanisms.

The drug doxorubicin (DOX) is an anthracycline antibiotic, and a leading chemotherapeutic agent with cytostatic effect used in the treatment of different types of human cancers [3]. Although doxorubicin is currently considered to be one of the most effective agents in the treatment of breast cancer, the drug use is limited by absolute maximum total doses that may be given and resistance to therapy frequently occurs, too. The molecular mechanism of action of doxorubicin interacts is based on DNA intercalation and inhibition of macromolecular biosynthesis [4]. The most diffused molecular mechanism involved in the killing of cancer cells is related to the activation of apoptotic pathway [4].

The main issue in the clinical use of doxorubicin is due to its heavy side effects on patients, especially for its intrinsic and cumulative dose-dependent cardiotoxicity [5-7]. To limit these side effects, doxorubicin can be encapsulated with drug nanocarriers that will allow targeted accumulation only on the cancer site and hence limit its dispersion into healthy tissues [8]. Superparamagnetic iron oxide nanoparticles (SPIONs), especially $Fe_3O_4$ nanoparticles, have been widely used in the field of biotechnology because of biocompatible, potentially non-cytotoxic, small size and interesting superparamagnetism properties [9]. Additionally, spherical magnetite NPs with diameters less than approximately 20 nm will exhibit superparamagnetic behavior, a property that is exploited to enhance contrast in magnetic resonance imaging (MRI) [9]. Multifunctional superparamagnetic iron oxide nanoparticles (SPIONs) for cancer theranostics can integrate chemotherapeutics, gene therapeutics, photothermal/photodynamic therapeutics, magnetic hyperthermia, fluorescent moieties and targeting moieties can be incorporated in the polymer coating through different strategies [8] (Figure 1).

Exploiting the interesting properties of the magnetic nanoparticles, the $Fe_3O_4$ nanoparticles were modified with polycarboxylic acid and loaded with the anticancer drug doxorubicin (DOX) for achieving a biocompatible, biodegradable and site-specific release of doxorubicin in the specific cancer site. Many types of tumors have a leaky neovasculature characterized by incomplete endothelial barrier and they possess poor lymphatic drainage. This effect, known as enhanced permeability and retention (EPR), allows nanosized carriers to easily reach their target site [10]. This study investigated both *in vitro* effect of induced magnetic field hyperthermia and targeted drug delivery of doxorubicin mediated by SPIONs on *in vitro* model of breast cancer. The novel dual targeted nanoparticles loaded with doxorubicin (DOX) and magnetic nanoparticles (SPIONs) can be considered a novel therapeutic tool to create personalized treatment in breast cancer patients. Great efforts have been made to develop drug carriers with the aim of providing predictable therapeutic response. Moreover, combination therapies have become promising strategies for clinical cancer treatment with synergistic effects.

The present study purposed to develop a new type of superparamagnetic nanocarrier for the intracellular co-delivery of doxorubicin (DOX) to the breast cancer cell line. A novel paramagnetic nanocomposite was designed to combine Rhodamine B and polycarboxylic SPIONs with doxorubicin to enhance anticancer activity.

Keeping in mind the novel properties of the magnetic nanoparticles, the $Fe_3O_4$ nanoparticles were modified with polycarboxylic acid and Rhodamine B and loaded with the anticancer drug doxorubicin (DOX) for achieving a biocompatible, biodegradable and site-specific carrier of anticancer drugs. The effect of these drug loaded nanocompounds was also evaluated on biochemical and morphological parameters of the human breast cancer cell lines of MCF-7 and MDA-MB-231. The *in vitro* cytocompatibility of the conjugate SPIONs-DOX and iron oxide nanoparticle-mediated hyperthermia was also evaluated on human mammary epithelial cell line (MCF-10A).

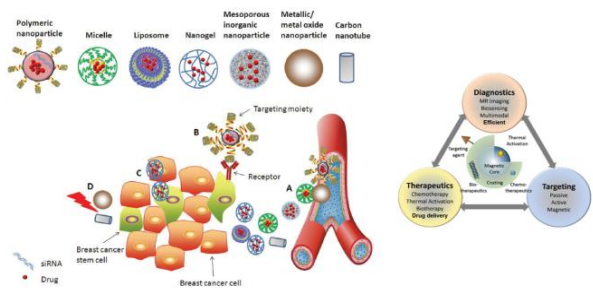

**Figure 1**. Various approaches explored to target breast cancer using nanomedicines.

**Breast cancer nanotheranostics**

The next generation cancer nanomedicine theranostics approach used for interventional oncology allows the diagnosis and treatment of cancer using targeted minimally invasive procedures performed under image guidance [18]. The current use of iron-oxide nanoparticles in breast cancer falls into four main groups: (1) imaging of primary and metastatic disease, (2) sentinel lymph node biopsy (SLNB), (3) drug delivery systems, and (4) magnetic hyperthermia [18]. The current evidence for the use of MNPs in these fields is mounting, and potential cutting-edge clinical applications, particularly with relevance to the fields of breast oncological surgery, are emerging [19]. Radiation therapy is a mainstay of treatment for patients with high grade gliomas, including glioblastoma. Radiation therapy in conjunction with surgery has been shown to prolong survival and, in the short term, improve cognitive function in patients with brain tumors. Over the longer term, however, radiation can cause fatigue and serious, permanent side effects, including radiation necrosis. Proton beam therapy, on the other hand, delivers very precise, very high doses of radiation to a tumor site, while sparing the surrounding healthy tissue [18, 19].

**Materials and methods**

**Synthesis of n(COOH) SPIONs**

SPIONs were synthesized via the co-precipitation method. Initially, $FeSO_4 \cdot 7H_2O$ and $FeCl_3$ (Sigma-Aldrich) were dissolved in water in 1:2 M ratios under nitrogen protection. The resulting dark orange solution was stirred for 30 min at 80 °C. An aqueous $NH_3$ solution (1.5 M) was then added dropwise to the above hot solution with stirring over a period of 15 min. Instant color change from dark orange to black was found to occur with particle formation. Stirring was continued for further 30 min followed by cooling to room temperature. Then 4 mL of an aqueous solution of citric acid (0.5 g/mL) was added to the above reaction mixture and reaction temperature was slowly raised up to 90 °C under reflux and reacted for 60 min with continuous stirring. The solvent was removed by magnetic decantation. Washing of the particles was done several times with water and then ethanol to make the iron dispersion free of any residual salts, which was used during the coprecipitation. The final supernatant was decanted magnetically to obtain the as-prepared superparamagnetic $Fe_3O_4$-$(COOH)n$ NPs.

**Surface coating of SPIONs with Rhodamine-B**

About 2 mg RhB was dissolve into 2 mL DMSO : $H_2O$ (1 : 1), then, a solution of EDC (8.7 mg, 10 mmol) and NHS (5.2 mg, 10 mmol) in MES buffer (2 mL) was added and incubated for 3 min. The obtained product was preserved in refrigerator at 4 °C. Afterwards 12 mg of the iron-oxide nanoparticles were introduced into 25 mL deionized water. After a brief sonication, solution of 5 mg of copper sulfate in 0.5 mL of water and 22 mg of sodium ascorbate in 1.0 mL of water were added, and the mixture was stirred for 30 min, then mixed with RhB and stirred in the room temperature all night. The final product was recovered by centrifugation and washed with PBS (pH = 7.0).

**Physicochemical Characterization of iron-oxide nanoparticles**

The morphology and size distribution of SPIONs were studied by transmission electron microscopy (TEM) and the hydrodynamic size and surface potential were measured through dynamic light scattering (DLS) and zeta potential, respectively. Furthermore, the crystalline phase of SPIONs was confirmed by X-ray diffraction (XRD, $\lambda$ = 0.15406 nm) and was detected by Fourier-transform infrared spectroscopy (FTIR). Cell internalization of SPIONs was assessed inside the MCF7 cells by ICP-MS after 2 h incubation with SPIONs.

**Experiment**

X-Ray powder diffraction (XRD) measurements were performed on a powder sample of the $Fe_3O_4$ nanoparticles using a Rigaku D/Ultima IV X-ray diffractometer (Rigaku, JAPAN), which was operated at 35 kV and 40 mA at a scan rate of 0.4 deg $s^{-1}$ and $2\theta$ ranges from 10° to 90° at room temperature using Cu-Ka radiation ($\lambda$ = 0.1542 nm). Transmission electron microscope (TEM) images were obtained at 200 kV by a JEM 2010 (JEOL, JAPAN) instrument. A drop of suspension of magnetite in ethanol (20 mg $mL^{-1}$) before and after modification was placed on a carbon-coated copper grid (Xinxin Bairui Corp, Beijing, China) and air dried. Fourier transform infrared (FTIR) spectra were recorded on a NEXLIS FTIR (Nicolet Corporation) and samples were dried at 90 C vacuum for at least 3 h prior to fabrication of the KBr pellet. In this context 4 mg of each sample was thoroughly mixed and crushed with 400 mg of KBr, and 70 mg of that mixture was used for pellet fabrication. Thirty-two scans of the region between 400 and 4000 $cm^{-1}$ were collected for each FTIR spectrum recorded.

**Dynamic light scattering (DLS) measurement of particle size distribution**

DLS was applied to measure size distribution of released DOX-SPIONs. Briefly, DOX-SPIONs nanoparticles were suspended in 200 µl of PBS (pH 7.4 or 6.6), followed by shaking at 300 rpm in a thermomixer (Eppendorf, Germany) at 37°C for 12 h. Supernatants containing released DOX-SPIONs nanoparticles were collected by centrifugation at 2,000 rpm for 4 min. SPIONs controls were prepared by the same procedure. DOX-SPIONs controls were prepared by direct dispersion of 5 µl DOX DMF solution into 2 mL PBS (pH 7.4 or 6.6). Particle size was measured with a Zetasizer Nano-ZS (Malvern Instruments Ltd, Worcestershire, UK) with measurements made using intensity average.

**Stability of SPIONs**

The stability of SPIONs was investigated by recording the change in turbidity in 50% FBS. 150 µl of SPIONs suspension (100 µg/ml) was added to 150 µl of FBS in 96-well plates and incubated for different times up to 72 h at 37 °C. After that, the absorbance of samples was measured at 405 nm. A solution of 5% glucose was employed as a negative control [9].

**Conjugation of DOX to the SPIONs**

For drug conjugation to modified SPIONs, different concentrations (50 and 100 µg/ml) of negatively charged polycarboxylic iron-oxide nanoparticles SPIONs were first sonicated with 0.5, 1, 5 and 50 µM concentrations of doxorubicin (DOX – Sigma-Aldrich) solution for 0.5 h and then stirred overnight at room temperature in the dark. DOX loading of modified SPIONs at an initial drug concentration of 50 µM was nearly saturating. The conjugation of doxorubicin to Rhodamine-B SPIONs was done by mixing 0.5, 1, 5 and 50 µM concentrations of doxorubicin (DOX) to 50 µg/ml RhB-SPIONs and then separating the DOX-RhB-SPIONs from precursors.
All the samples were centrifuged at 18 000 x g for 1 h. The DOX concentration of all the samples was measured using a standard DOX concentration curve, generated with a UV-Vis spectrophotometer (Cary 60 UV-Vis – Agilent) at the wavelength of 233 nm. The drug conjugation to the n(COOH)-SPIONs has been illustrated in Figure 2.

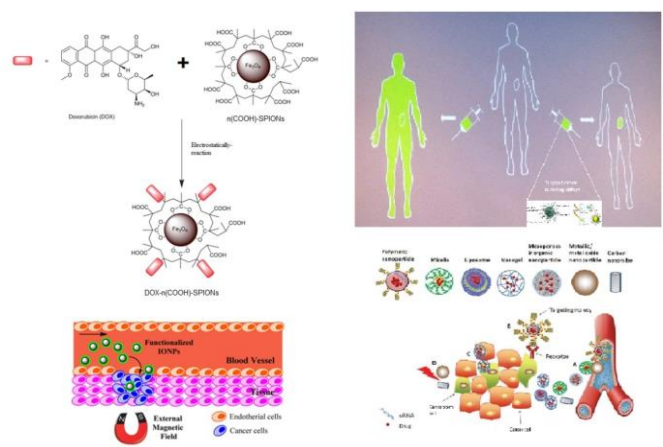

**Figure 2.** Conjugation of doxorubicin (DOX) to the n(COOH)-SPIONs.

**Cell cultures**

The human breast cancer cell lines MCF-7 (ATCC - HTB-22) and MDA-MB-231 (ATCC - HTB-26) cells were obtained from the American Type Culture Collection and cultured in DMEM, containing 1% (v/v) penicillin-streptomycin and 10% (v/v) heat-inactivated FBS, respectively. FBS and penicillin-streptomycin were obtained from GIBCO - Life Technologies, Carlsbad, CA, USA.
Human mammary epithelial cells (MCF 10A, ATCC, CRL-10317) was cultivated in ™ Mammary Epithelial Cell Growth Medium (MEGM, Lonza/Clonetics) fortified with 2 mM L-glutamine, 100 ng/ml cholera toxin, 10% fetal bovine serum (FBS, Gibco, Thermo-Fisher, Pittsburg, PA, US) and 1% antibiotics/antimycotics (penicillin/streptomycin/gentamycin, Gibco) (complete medium). Cells were incubated in a humidified 5% $CO_2$ atmosphere at 37°C and passaged twice weekly at a 1:4-5 concentration.

**Magnetic hyperthermia in breast cancer and normal cells**

Breast cancer cells were exposed to hyperthermia conditions to induce mediated apoptosis. For hyperthermia treatment, at 24 h after seeding 5000 cells/well in a 96-well plate, cells were incubated with either medium, DOX-SPIONs in a concentration of 50 µg Fe/ml or the equivalent molar amount of free DOX at 46°C for 30 minutes in contact with a magnet, corresponding to a temperature dosage of 90 cumulative equivalent minutes at 43°C, or else cells was left in the incubator at 37°C. At 48 h after hyperthermia, cells viability was tested. The in vitro efficacy was also tested without hyperthermia treatment. Immediately afterward, the viability of cells was assessed by MTT assay.

**Leaching toxicity tests of MNPs**

For leaching toxicity tests the SPIONs were sterilized with UV and MEGM culture medium containing 10% calf serum was then added to it in a sterile closed vessel (final concentration 0.1 g/mL). Leaching conditions were 37°C over 72 hours. The samples were then centrifuged at 2500 rpm for 5 minutes, the supernatant was removed by suction and filtered through a microporous filter (VWR International, U.S.) to provide a 100% leach solution. During the test, the MEGM culture medium was diluted to the required concentration. MCF-10A cells were removed during their logarithmic growth phase and digested in the cell suspension; the cell concentration was next adjusted to $4 \times 10^3$/mL and inoculated on a 96-well culture plate at 200 µl/well. The sample was cultured in an incubator at 37°C under saturated humidity and 5% $CO_2$ conditions. The primary solution was discarded after 24 hours, at which time the leaching solution was added until the final concentration was 100%, 75%, 50%, and 25%. The MEGM culture inoculum was used as control. Each group comprising four wells was cultured for 72 hours, and 20 µl of MTT was then added into each well and incubated for 4 hours. The absorbance value was measured at 570 nm using a spectrophotometer. The relative cell growth rate was calculated as follows: relative growth rate % = OD (Optical Density) mean value of test group/OD mean value of negative control group × 100%.

**Cell viability and apoptosis analyses**

Viability was determined by MTT assay. Data obtained were used to calculate the $IC_{50}$ and $IC_{90}$ values for each cell line, administered with DOX-n(COOH)-SPIONs. For all experiments, $3 \times 10^6$ cells from each cell line were plated into 150 mm tissue culture dishes, 24 h prior to the nano-composite treatment. In order to determine the corresponding $IC_{50}$ and $IC_{90}$ values, the cells were separately treated with the as-prepared bare, coated and drug-loaded nano-composites for 24, 48, 72 and 96 h at 37 °C. Plates were washed thrice with PBS and incubated in 10 mL of fresh medium for an additional 48 h before the final cell collection. All the cells were collected by sedimenting at 2000 x g for 5 min. For the 5-dimethylthiazol-2-yl-2,5-diphenyltetrazolium bromide (MTT) assay, MCF-7 and MDA-MB-231 cells were plated in 96-well plates and exposed to n(COOH)-SPIONs, RhB-SPIONs and DOX-n(COOH)-SPIONs at a concentration of 50 µg/mL for 24, 48, 72 and 96 h. After the culture period, the medium was removed from each well and replaced with fresh medium. Cellular apoptosis was detected by Annexin-V-FLUOS staining and studied with the help of flow cytometry.

**Direct cytotoxicity evaluation of SPIONs and DOX-SPIONs on cancer cells**

Direct cytotoxicity of SPIONs on breast cancer cells was evaluated following the ISO standard 10993-5:2009 on Biological Evaluation of Medical Devices instructions [3]. Cells were seeded in 24-well plates (1.6 x $10^4$ cells per well) in complete DMEM medium and incubated for 24, 48 and 72 h at 37°C in 5% $CO_2$. Different combinations of hyperthermia and targeted drug delivery of doxorubicin were tested on breast cancer cells to observe their efficacy. Fresh medium without MNPs was used as control. Cell viability was evaluated after 24, 48 and 72 hours by the (3-(4,5-Dimethylthiazol-2-yl)-2,5-diphenyltetrazolium bromide colorimetric assay (MTT, Sigma). Briefly, 100 µl of MTT solution (3 mg/ml in PBS) were added to each specimen and incubated for 4 hours in the dark at 37°C; formazan crystals were then dissolved in 100 µl of dimethyl sulphoxyhde (DMSO, Sigma- Aldrich) and 50 µl were collected and centrifuged to remove any debris. Supernatant optical density (o.d.) was evaluated at 570 nm. The mean optical densities obtained from control specimens were taken as 100% viability. Cell viability was calculated as follow: (experimental o.d. / mean control o.d.)*100. Experiments were performed with four replicates at each experimental time.

**Cell uptake analysis**

Visual determination of cell uptake and incorporation of Rhodamine B-SPIONs were performed using optical microscopy with fluorescent mode. For comparison with in vitro battery tests used in this work, we tracked the efficiency of labeling and cell morphology after 24 hours of incubation of cells with RhB-SPIONs. All experiments were performed on two types of cell lines (MCF7 and MDA-MB-231).

**$IC_{50}$ and $IC_{90}$ evaluation of SPIONs**

Viability was determined by MTT assay. Data obtained were used to calculate the $IC_{50}$ and $IC_{90}$ values for each cell line, administered with DOX-n(COOH)-SPIONs and DOX-Rhodamine B-SPIONs. For all experiments, 3 x $10^6$ cells from each cell line were plated into 150 mm tissue culture dishes, 24 h prior to the nano-composite treatment.

In order to determine the corresponding IC$_{50}$ and IC$_{90}$ values, the cells were separately treated with the as-prepared bare, coated and drug-loaded nano-composites for 24, 48, 72 and 96 h at 37 °C. Plates were washed thrice with PBS and incubated in 10 mL of fresh medium for an additional 48 h before the final cell collection. All the cells were collected by sedimenting at 2000 x g for 5 min.

**Statistical analysis**

All statistical analyses were performed using the IBM Statistical Package for Social Sciences v. 20 (SPSS-IBM, Chicago, MI, US). All data were expressed as mean ± SD and one-way ANOVA was used for statistical analysis. P values < 0.05 were considered statistically significant.

# Results

X-ray diffractometry of magnetic nanoparticles was used to identify the crystal structure and estimate the crystallite size of as-prepared nanoparticles (Fig. 3 A-B). The lattice constant *a* was measured to be 8.310 Å, which was compared with the lattice parameter for the magnetite of 8.39 Å [12]. As shown in Fig. 3a, the peaks indexed as planes (220), (311), (400), (422), (511) and (440) correspond to a cubic unit cell, characteristic of a cubic spinel structure [13]. In addition, the strongest reflection from the (311) plane is also the characteristic of such phase. Therefore, it is confirmed that the crystalline structure of obtained magnetite nanoparticles is an inverse spinel type oxide, in accord with that of standard data (JCPDS 72-2303).

The crystal structure of $Fe_3O_4$ is very similar to that of $\gamma$-$Fe_2O_3$, and hence it is very difficult to distinguish them from each other by XRD. Thus, XPS was used to further analyze the composition signals. As shown in Fig. 3b, the XPS wide-scan spectrum of the $Fe_3O_4$ nanoparticles is dominated by the Fe signal at the binding energy of about 720 eV and the O signal at the binding energy of about 550 eV. The C1s core-level spectrum in Fig. 3b with binding energy at about 280 eV is attributed to the C=O–O species of the citrate acid grafted on the nanoparticle surface.

**FTIR spectra**

The FTIR spectra of $Fe_3O_4$ and functionalized RhB-$Fe_3O_4$ nanoparticles are shown in Fig. 3C. The absorption that is recorded at 1590 $cm^{-1}$ reveals the presence of the Rhodamine-B group, which also suggests the successful modification of RhB (Fig. 3C). The spectrum for $Fe_3O_4$ nanoparticles is shown in Fig. 5b. As shown for RhB-$Fe_3O_4$ (Fig. 3C), the characteristic bands corresponding to Rhodamine group indicates the successful proceeding of reaction. It also suggests that the functionalization reaction is performed with a high yield [15].

**Fluorescence spectra of Rhodamine B iron-oxide nanoparticles**

To examine whether the RhB-$Fe_3O_4$ nanoparticles are photostable, it was performed a photobleaching experiment in which samples were repeatedly illuminated and the corresponding fluorescence intensities were measured [16]. It was observed that the fluorescence intensity of the RhB-$Fe_3O_4$ nanoparticles, during about 48 h illumination, remains almost unaltered. However, as a control experiment, 96% decrease in fluorescence intensity was observed due to the photobleaching of free Rhodamine-B (Fig. 3 D-E). The preserved fluorescent in the RhB-$Fe_3O_4$ nanoparticles can be explained by the formation of stable reaction.

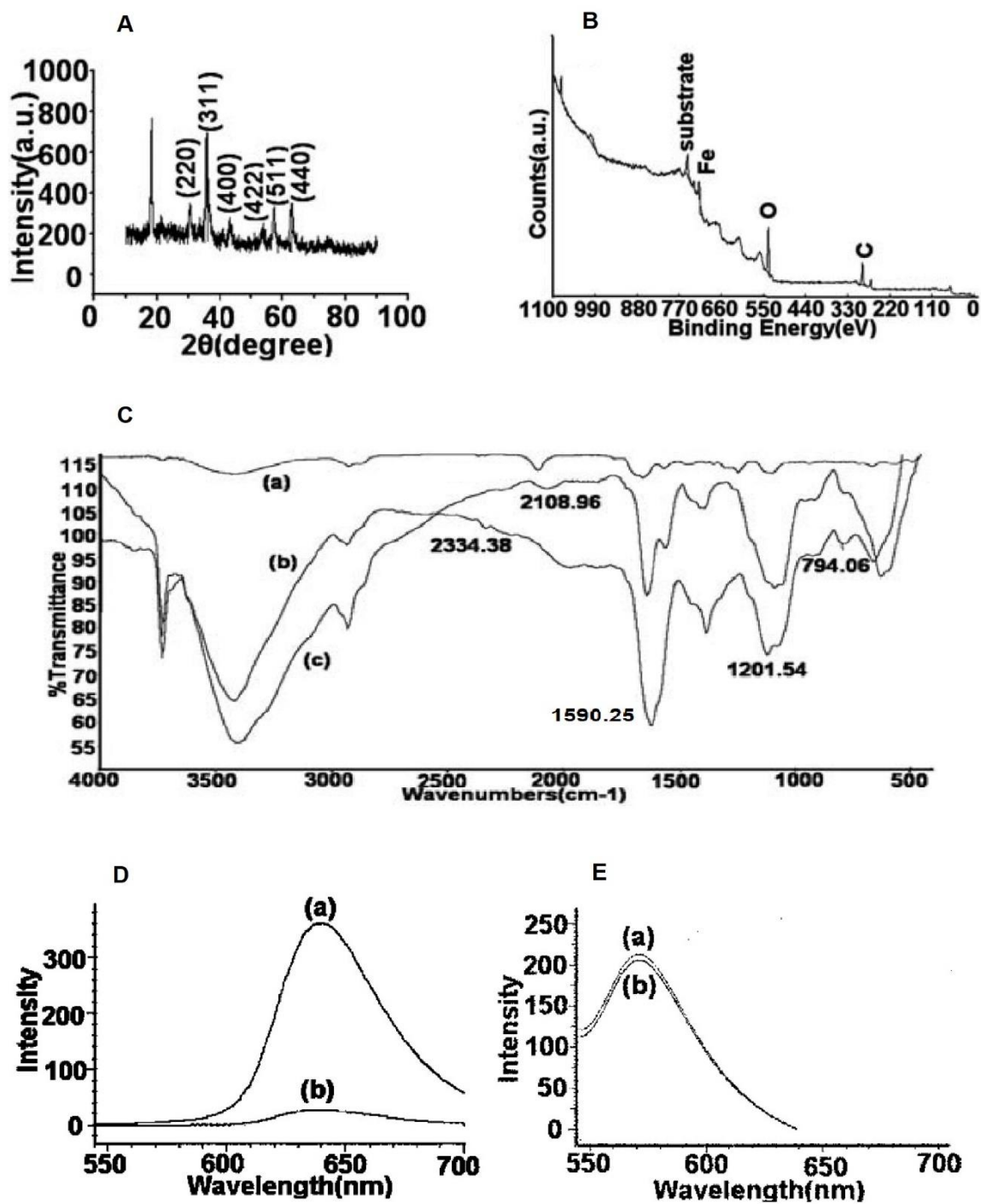

**Figure 3.** (**A**) XRD patterns of $Fe_3O_4$, (**B**) XPS wide-scan spectra of the as-synthesized $Fe_3O_4$ nanoparticles; (**C**) FTIR of Rhodamine-B (RhB) (a), $Fe_3O_4$ (b), RhB-$Fe_3O_4$ (c); (**D**) Fluorescence spectra of RhB and (**E**) RhB-$Fe_3O_4$ in PBS in the beginning (a), and after 48 hours (b).

**TEM physicochemical characterization of SPIONs**

The TEM images of the $Fe_3O_4$ and $RhB-Fe_3O_4$ nanoparticles are shown in Fig. 4. The bare $Fe_3O_4$ nanoparticles (Fig. 4) appear aggregated, which is probably a consequence of the removal of the dispersing phase during sample preparation [14].

The TEM image (Fig. 4) of $RhB-Fe_3O_4$ shows that the core of $Fe_3O_4$ can be observed. It is very clear that the nanoparticles have a discrete core/shell structure. Meanwhile, the size distribution of the $Fe_3O_4$ nanoparticles (Fig. 4) is very narrow over the wide range of the TEM grid area, and the $RhB-Fe_3O_4$ nanoparticles exhibit considerable dispersion. From TEM analysis, the average diameter of the particles was calculated to be 15-20 nm, which is consistent with the XRD analysis.

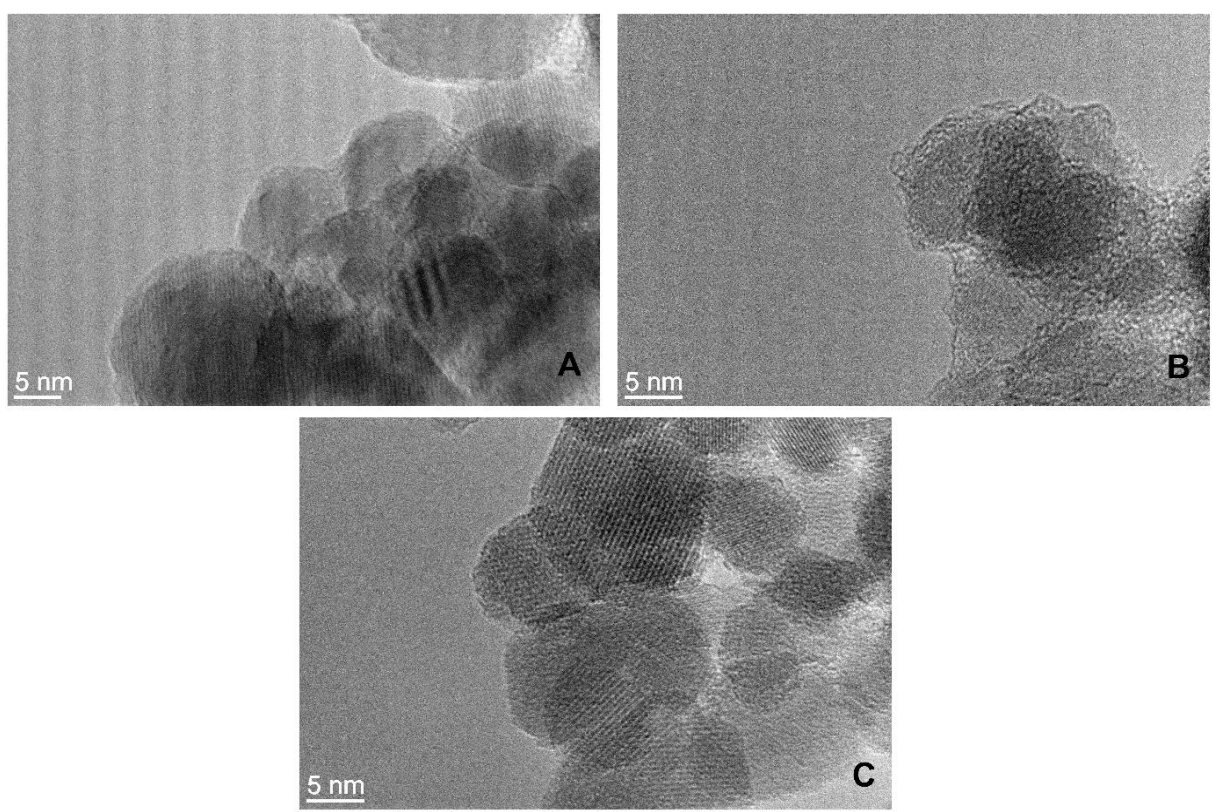

**Figure 4.** TEM of Fe3O4 (a) and $RhB-Fe_3O_4$ (b and c).

**Dynamic light scattering analysis**

Size distribution of the synthesized n(COOH)-SPIONs was investigated in an aqueous solution with the help of dynamic light scattering (DLS) spectrometer. As shown in the Table 1 the mean diameter of the n(COOH)-SPIONs was recorded to be about 92.0 nm and DOX-loaded polycarboxylic-coated SPIONs was 154.4 nm. The size distribution of the bare SPIONs was also determined and recorded to be about 30 nm. The average hydrodynamic diameter increases with increased amount of doxorubicin from 92 nm to 254.7 nm. The mean diameter of the $RhB-Fe_3O_4$ nanoparticles was recorded to be about 74.0 nm and DOX-loaded Rhodamine B-coated SPIONs was 145.7 nm.

**n(COOH)-SPIONs**

| Combination SPIONs/Doxorubicin | Average Size (DLS) | Zeta Potential |
|---|---|---|
| 50 µg/ml / 0 µM | 92 nm | - 45 mV |
| 50 µg/ml / 0.5 µM | 137.2 nm | - 37.4 mV |
| 50 µg/ml / 1 µM | 109.9 nm | - 49.9 mV |
| 50 µg/ml / 5 µM | 154.4 nm | - 34.1 mV |
| 50 µg/ml / 50 µM | 254.7 nm | - 41.7 mV |

**Rhodamine B-SPIONs**

| Combination SPIONs/Doxorubicin | Average Size (DLS) | Zeta Potential |
|---|---|---|
| 50 µg/ml / 0 µM | 74 nm | - 30 mV |
| 50 µg/ml / 0.5 µM | 103.4 nm | - 25 mV |
| 50 µg/ml / 1 µM | 124.6 nm | - 31.9 mV |
| 50 µg/ml / 5 µM | 145.7 nm | - 25.9 mV |
| 50 µg/ml / 50 µM | 201.6 nm | - 39.3 mV |

**Table 1.** Dynamic light scattering (DLS) spectrometer of DOX-n(COOH)-SPIONs and DOX-Rhodamine B-SPIONs

**Colloidal stability and hydrodynamic behaviour of SPIONs**

The colloidal stability of n(COOH) and RhB-SPIONs was investigated before cell viability evaluation. The hydrodynamic size was measured in a narrow range by DLS evaluation after 60 days, as shown in Fig. 5. There was no sign of aggregation and flocculation or subsidence after 60 days of storage (Fig. 5).

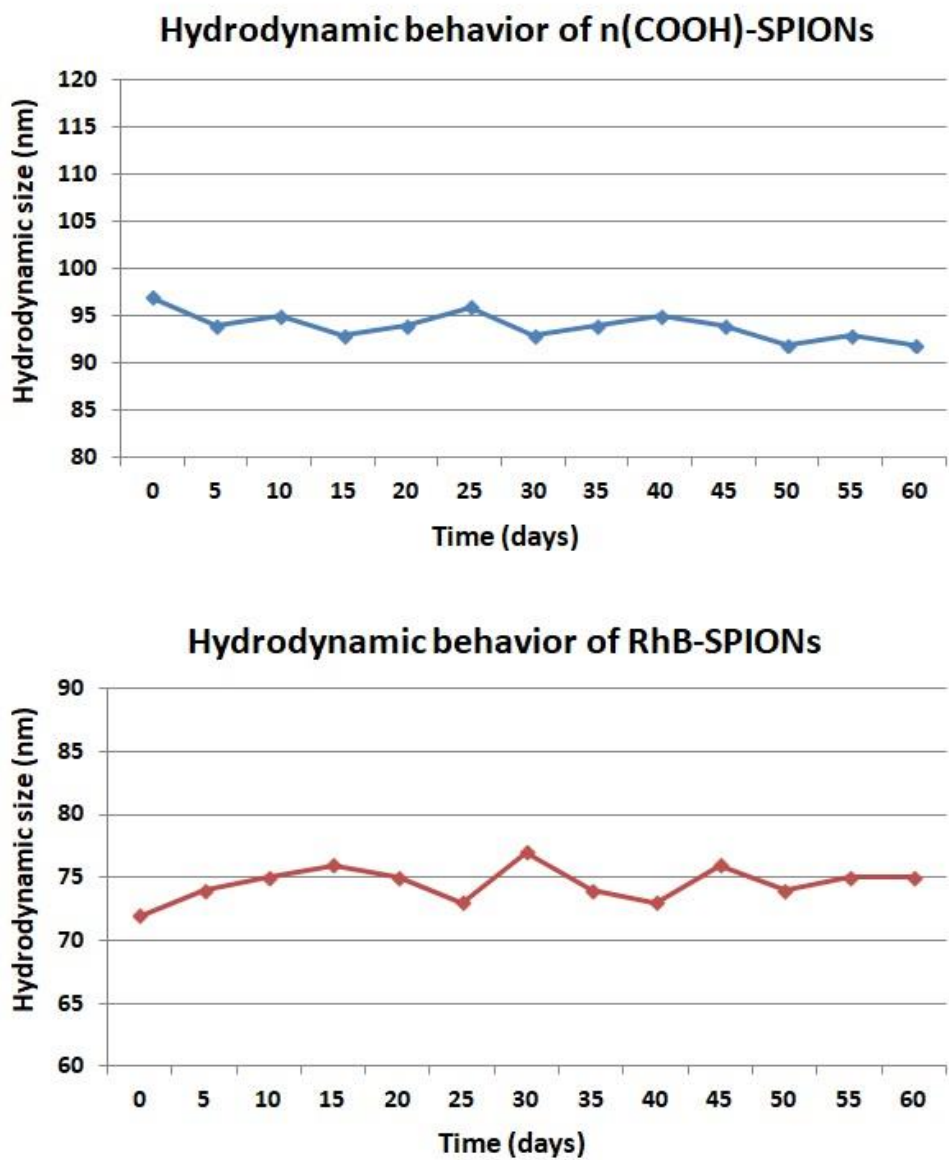

**Figure 5.** Hydrodynamic sizes of n(COOH) and RhB-SPIONs in aqueous solution as a function of storage time.

**Internalization of SPIONs into cells**

The results of ICP-MS showed that higher amount of iron (45.8 ± 1.8 pg) was taken up in higher SPIONs concentration (100 µg/ml) (concentration-dependent). After identical sample preparation in the control, 10.1 ± 0.9 pg iron per cell could be detected. In the case of 50, 100 µg/ml of SPIONs, the numbers were 24.9 and 45.8 pg/cell, respectively for n(COOH)-SPIONs, and 21.8 and 38.7 pg/cell for RhB-SPIONs (Fig. 6).

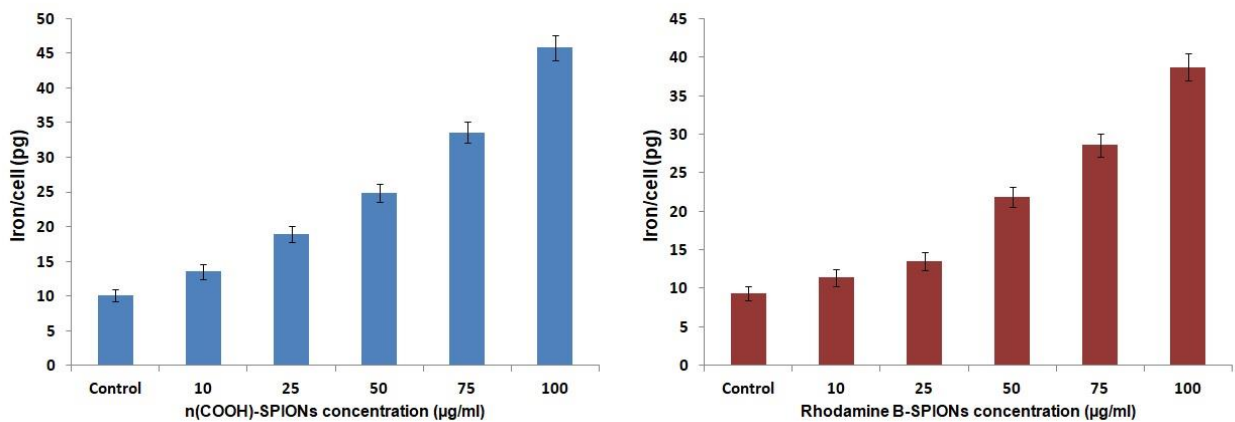

**Figure 6.** ICP-MS of different concentrations of n(COOH) and Rhodamine B SPIONs.

**Material toxicity testing of magnetic nanoparticles leach solution**

The values of cell viability revealed no significant difference in MCF-10A cell growth compared with that in the control group after n(COOH)-$Fe_3O_4$ and Rhodaminie B $Fe_3O_4$ nanoparticle leaching solution (100%, 75%, 50%, and 25%) was added. The relative growth rate is shown in Table 2. The results indicate that the 100% leaching solution of n(COOH) and RhB $Fe_3O_4$ nanoparticles is cytocompatible toward MCF-10A cells. Considering the relative growth rate, the eluates solutions of n(COOH)$Fe_3O_4$ and RhB-$Fe_3O_4$ nanoparticles showed a cell viability on MCF-10A cells comparable to control.

| n(COOH)-$Fe_3O_4$ nanoparticles | | | RhB-$Fe_3O_4$ nanoparticles | | |
|---|---|---|---|---|---|
| **Groups** | **OD** | **Relative growth rate (%)** | **Groups** | **OD** | **Relative growth rate (%)** |
| **Control** | 0.289 ± 0.005 | **100** | **Control** | 0.289 ± 0.005 | **100** |
| **25% extract liquid** | 0,277 ± 0.004 | **95.9** | **25% extract liquid** | 0.280 ± 0.008 | **97.1** |
| **50% extract liquid** | 0.271 ± 0.006 | **93.8** | **50% extract liquid** | 0.273 ± 0.005 | **94.6** |
| **75% extract liquid** | 0.261 ± 0.008 | **90.6** | **75% extract liquid** | 0.266 ± 0.007 | **92.1** |
| **100% extract liquid** | 0.254 ± 0.006 | **87.8** | **100% extract liquid** | 0.257 ± 0.006 | **88.9** |

**Table 2.** Leaching extract solutions for n(COOH)-$Fe_3O_4$ and Rhodamine-B $Fe_3O_4$ nanoparticles.

**Direct contact cytotoxicity**

The same conditions were verified on MCF-10A cells. For MCF-10A cells the cell viability after 24 hours (MTT assay) range between 80% and 88% for n(COOH)-$Fe_3O_4$ and RhB-$Fe_3O_4$ (Fig. 7). DMSO (10% v/v) was used like negative control. The cell viability after 48 hours (MTT assay) range between 83% and 91% for n(COOH)-$Fe_3O_4$ and RhB-$Fe_3O_4$ (Fig. 7). The cell viability after 72 hours (MTT assay) range between 82% and 92% for n(COOH)-$Fe_3O_4$ and RhB-$Fe_3O_4$ (Fig. 7). RhB-$Fe_3O_4$ and n(COOH)-$Fe_3O_4$ SPIONs did not affect the viability of mammary epithelial cells in all the experimental conditions. These data of cell viability showed cytocompatibility and stability of the layer of polycarboxylic acid and Rhodamine B comparable to control with a selectivity for cancer cells compared to non cancerous cells.

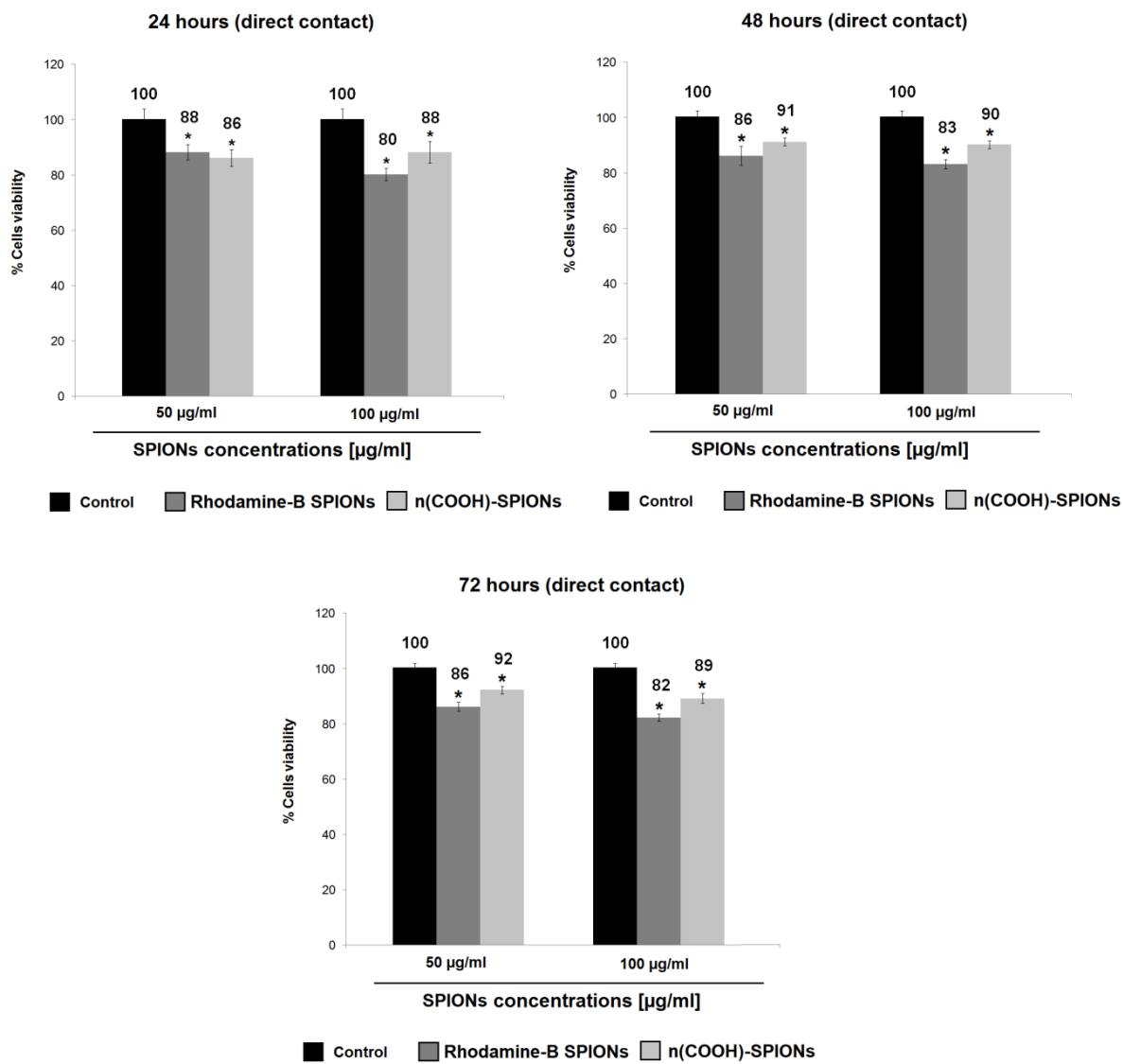

**Figure 7.** Direct contact cytotoxicity evaluation of n(COOH)-SPIONs and RhB SPIONs (50 and 100 µg/mL) on MCF-10A cells at different time points. *P < 0.05 compared with control samples.

### Direct cytotoxicity evaluation of SPIONs and DOX-SPIONs on cancer cells

The cytotoxicity of DOX loaded n(COOH)-SPIONs was compared with free drug DOX in the absence and presence of magnetic field, also the probable cytotoxicity of the targeted and non-targeted blank nanoparticles was checked. Cell viability assay showed that in MCF-7 and MDA-MB-231 cells the cell survival percentage was decreased significantly (p<0.05) in targeted nanoparticles group compared to free DOX and naked magnetic nanoparticles in most drug concentrations both in absence and presence of magnetic field without toxic effects on normal mammary epithelial cells (MCF-10A). These data indicate that DOX-CA-SPIONs and the presence of magnetic hyperthermia led to increasing cytotoxic effects of the nanoparticles on breast cancer cells.

A selective efficacy was observed for DOX-SPIONs on breast cancer cells. The results demonstrated the potential of the DOX-n(COOH) SPIONs to achieve dual tumor targeting by magnetic field-guided in breast cancer cells, and exploit the incredible possibilities to kill in a target way the cancer cells.

Drug-loading efficiency of doxorubicin showed better results for the n(COOH)-SPIONs as compared to the bare ones. For hyperthermia treatment, cells were exposed to hypertermic conditions for 30 minutes, corresponding to a temperature dosage of 90 cumulative equivalent minutes, or else cells will be left in the incubator at 37°C. At 24, 28, 72 h after hyperthermia, cells viability was tested. The in vitro efficacy was also tested without hyperthermia treatment. The incubation of MCF-7 and MDA-MB-231 human breast cancer cells with doxorubicin-loaded and doxorubicin-loaded polycarboxylic acid-coated SPIONs, for 24, 48, 72 h, showed significant $IC_{50}$ and $IC_{90}$, respectively, after 72 h of incubation (Figure 8 and 9). While, 90% and 93% growth inhibition were seen in MCF-7 and MDA-MB-231 cells after the 72-h exposure to the doxorubicin n(COOH)-SPIONs ($p < 0.05$), any type of cytotoxic effect was observed on control MCF-10A cells (ATCC, mammary epithelial cells) (Figure 10). These data indicate that DOX-n(COOH)-SPIONs and the presence of magnetic hyperthermia led to increasing cytotoxic effects of the nanoparticles on breast cancer cells like a very promising moonshot for cancer therapy.

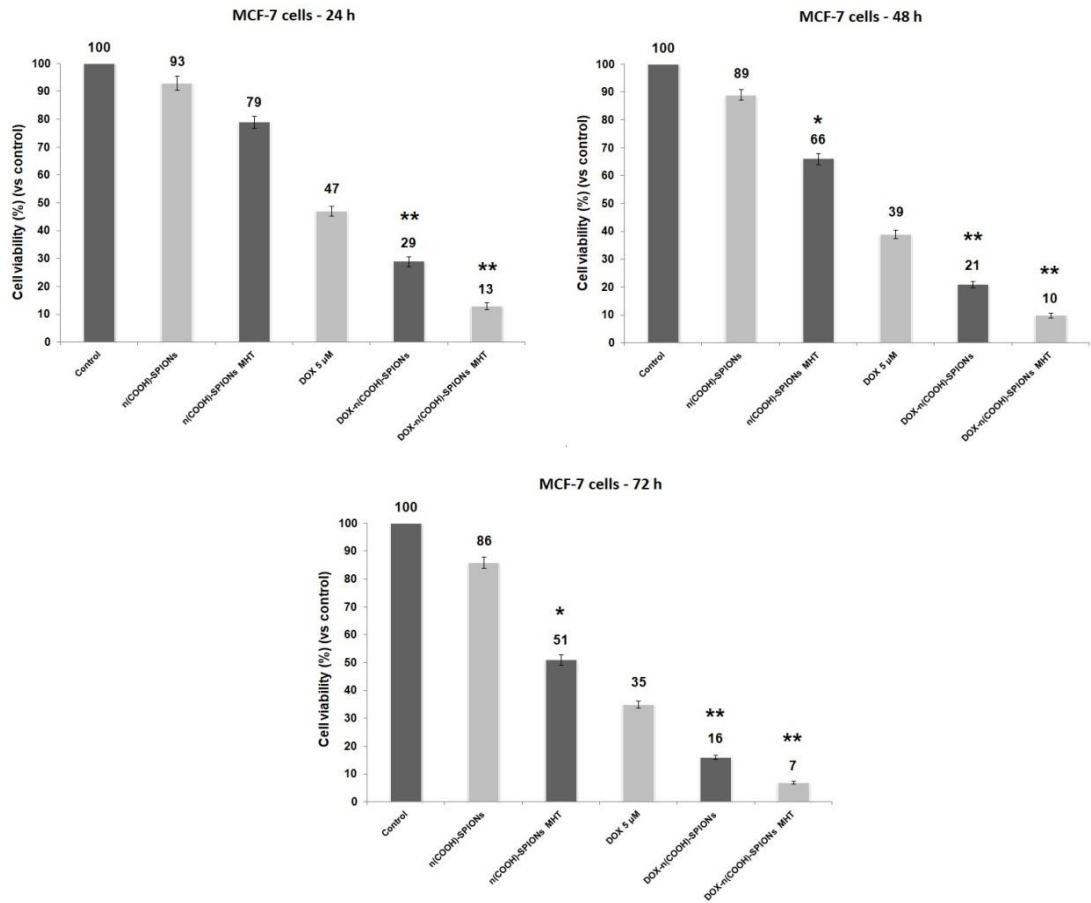

**Figure 8.** Direct contact cytotoxicity evaluation of n(COOH)-SPIONs (50 µg/mL) conjugated to Doxorubicin (5 µM) (DOX) and with induced magnetic hyperthermia using breast cancer cells (MCF-7 cells) at different time points. Data are shown as the mean ± standard error of the mean (n = 4). *P < 0.05 compared with control samples. **P < 0.05 compared to control samples and DOX 5 µM.

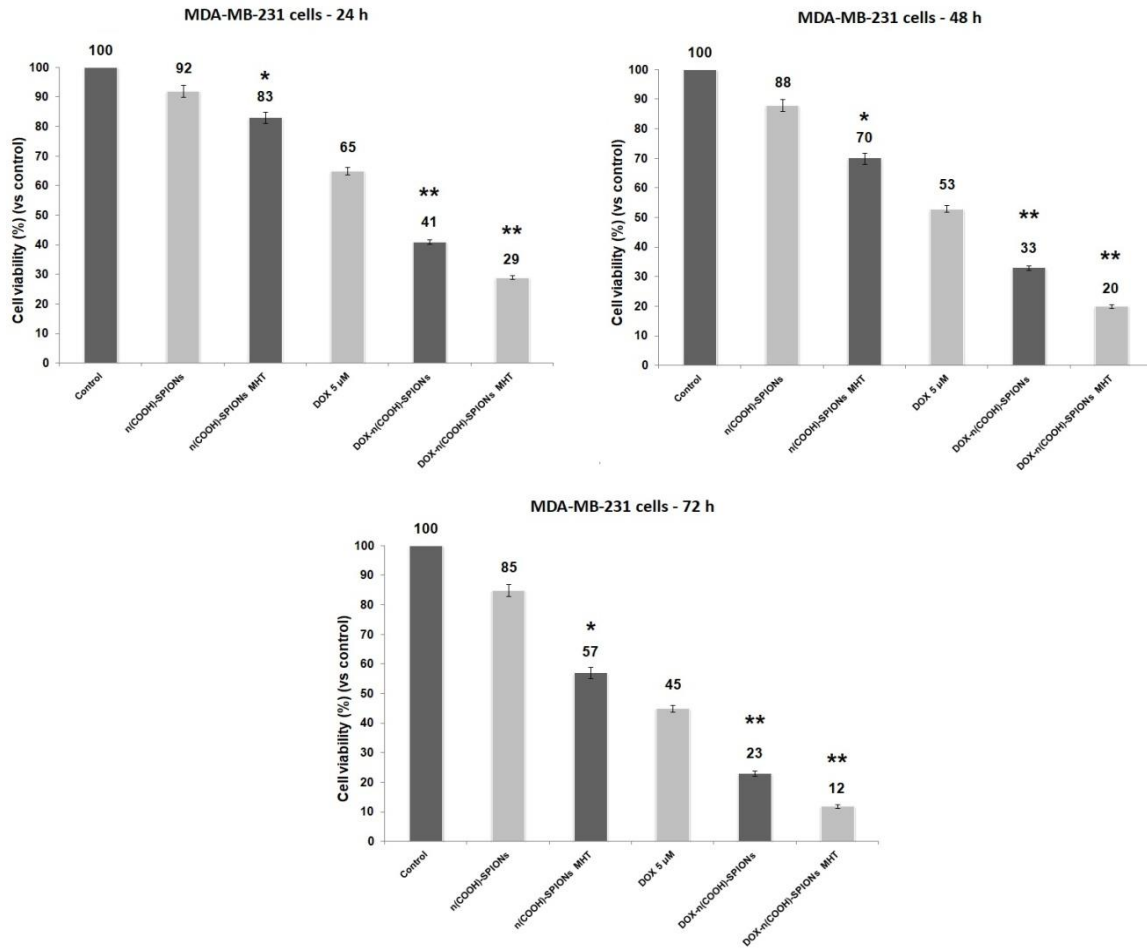

**Figure 9.** Direct contact cytotoxicity evaluation of n(COOH)-SPIONs (50 μg/mL) conjugated to Doxorubicin (5 μM) (DOX) and with induced magnetic hyperthermia using breast cancer cells (MDA-MB-231 cells) at different time points. Data are shown as the mean ± standard error of the mean (n = 4). *P < 0.05 compared with control samples. **P < 0.05 compared to control samples and DOX 5 μM.

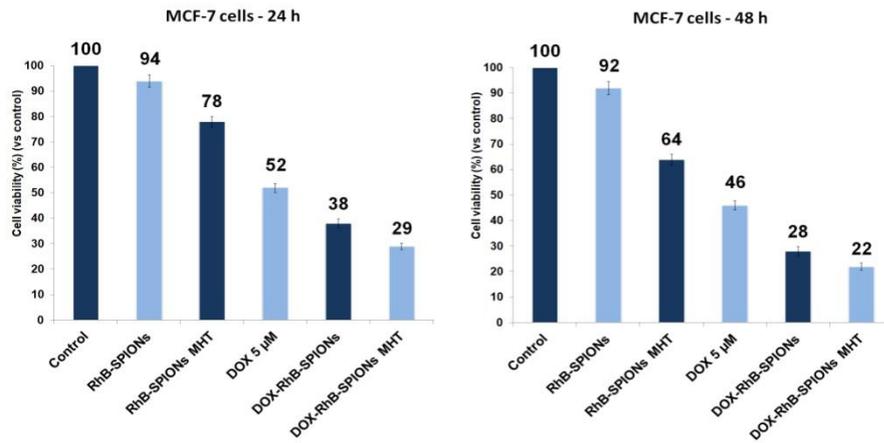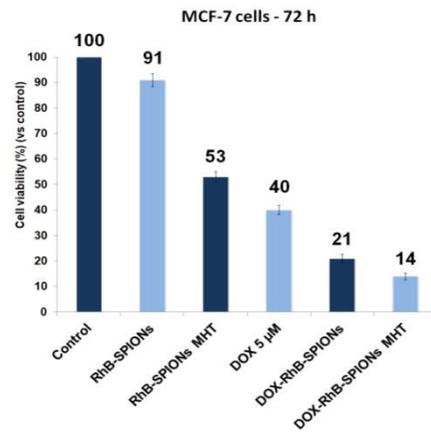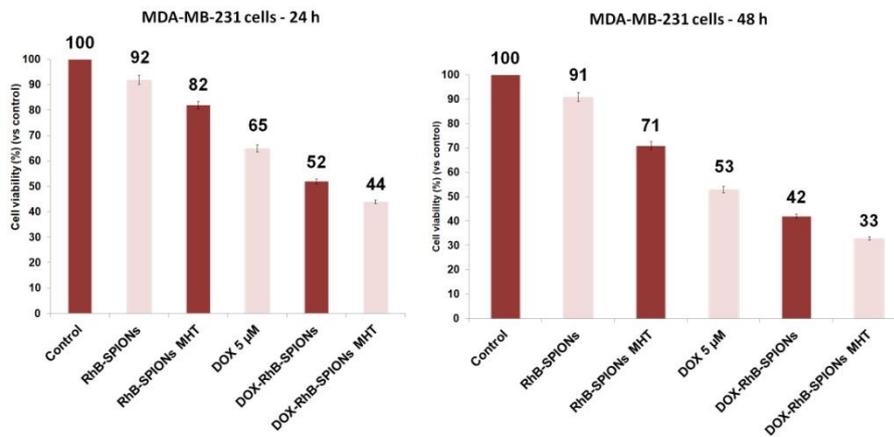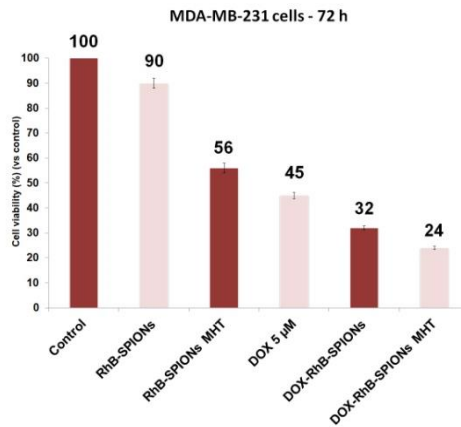

**Figure 10.** Direct contact cytotoxicity evaluation of RhB-SPIONs (50 µg/mL) conjugated to Doxorubicin (5 µM) (DOX) and with induced magnetic hyperthermia using breast cancer cells (MCF-7 and MDA-MB-231 cells) at different time points. Data are shown as the mean ± standard error of the mean (n = 4). *P < 0.05 compared with control samples. **P < 0.05 compared to control samples and DOX 5 µM.

**Apoptosis evaluation**

The results revealed a dose-dependent induction of early or late apoptotic cell death in the breast cancer cell lines (Figure 11). Compared to MDA-MB-231 cells, the MCF-7 cells showed more cell death. MCF-7 cells showed 91% apoptosis after the administration of DOX-loaded n(COOH)-SPIONs (5 µM; p < 0.05). While, 87.1% apoptosis was seen in the MDA-MB-231 cells under same conditions (p < 0.05), which indicates that carboxylic acid-functionalized SPIONs could enhance the DOX-induced apoptosis in both of the human breast cancer cell lines. For the cell death onset analysis, MCF-7 and MDA-MB-231 cells were treated separately with DOX, SPIONs, SPIONs MHT and DOX-n(COOH)-SPIONs AMF for 24 h of time. Apoptotic cell death was analyzed by staining the cells with Annexin-V-FLUOS staining kit and analyzed by flow cytometry. The results revealed a dose-dependent induction of early or late apoptotic cell death in these two cell lines (Figure 9). Compared to MDA-MB-231 cells, the MCF-7 cells showed more cell death. MCF-7 cells showed 91% apoptosis after the administration of DOX-loaded SPIONs (5 µM; p < 0.05). While, 87.1% apoptosis was seen in MDA-MB-231 cells under same conditions (p < 0.05), which indicates that hydroxyl-modified SPIONs could enhance the DOX-induced apoptosis in both of the human breast cancer cell lines.

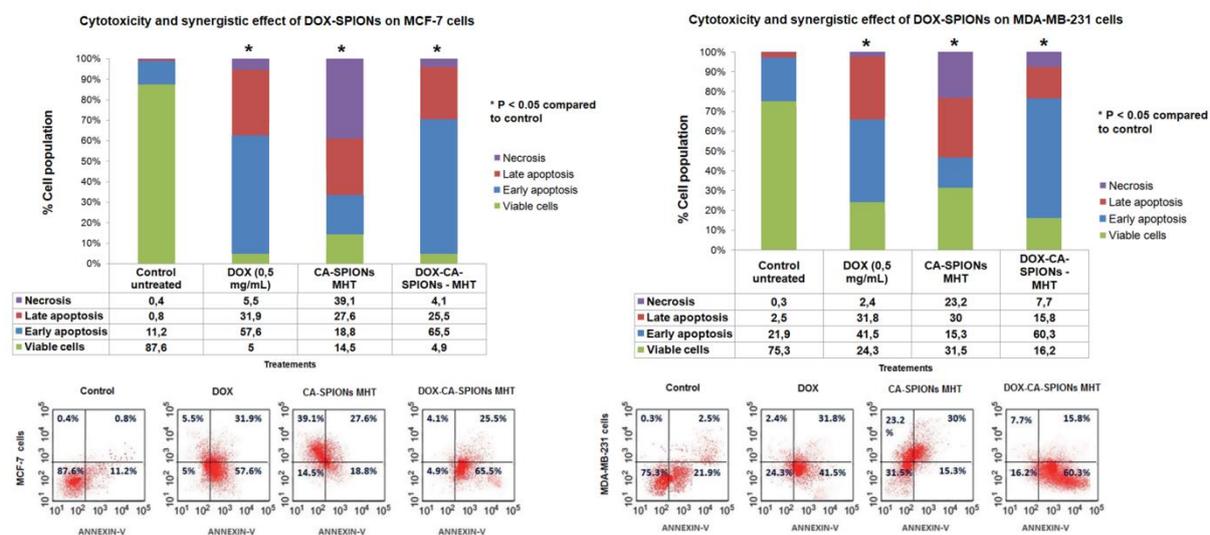

**Figure 11.** Doxorubicin (DOX)-loaded CA-SPIONs system induces apoptosis in MCF-7 and MDA-MB-231 cells. Both cancer cell lines were treated separately with DOX, SPIONs, CA-

SPIONs, and DOX-CA-SPIONs for 24 h. Apoptotic cell death was detected by staining the cells with Annexin-V/PI kit and analyzed by flow cytometry.

**Intracellular internalization of Rhodamine-B SPIONs**

The Rhodamine-B conjugated iron-oxide nanoparticles were internalized by both MCF-7 and MDA-MB-231 cells as detected by fluorescence microscopy (Figure 10). The image in Figure 12 represents the intracellular uptake of nanoparticles bound to doxorubicin to kill more specifically breast cancer cells. The co-localization of these cells involves the bases of long-time studies and successful internalization

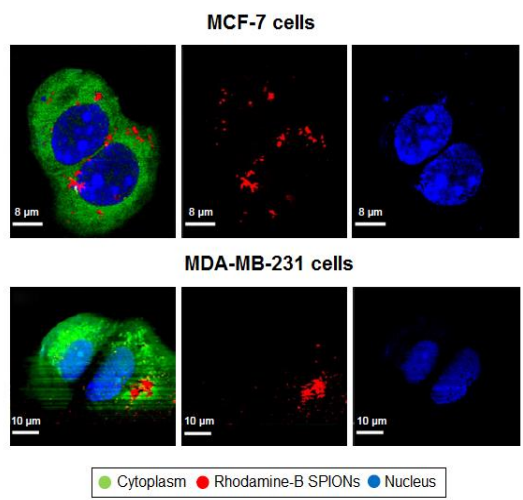

**Figure 12.** Intracellular Rhodamine B-conjugated iron-oxide nanoparticle (SPIONs) location by fluorescence microscopy.

**Notes:** Images of different components of MCF7 and MDA-MB-231 cells exposed to DOX-CA-SPIONs over 24 hours. The color-coded images on the left represent the overlapping of cytoplasm (green), nuclei (blue), and DOX-CA-SPIONs (red).

**TEM characterization of SPIONs uptake**

The effects of SPIONs and SPIONs-DOX on MCF-7 cells were analysed, as well as on their proliferation, by the MTT assay, which were measured in terms of relative viability. Conjugated iron-oxide nanoparticles were sparsely scattered throughout the cytoplasm and mostly accumulated in membrane structures.

Following 24 hours of incubation with nanoparticles, it's possible to observe the internalization of particles in the cytoplasm of MCF-7 cells (Figure 13). SPIONs mostly accumulated in the cytoplasm in the endosome-like structures surrounding the nucleus.

When the cells were analyzed by TEM, we could observe the presence of SPION-DOX conjugates included as electron-dense particles in membrane structures within the cytoplasm (Figure 14a). Transmission electron microscopy to characterize the subcellular localization demonstrated that SPIONs complexes were colocalized in endosomes (Figure 14b and c). This proposes a mechanism of receptor-mediated endocytosis of SPIONs-DOX conjugates by MCF-7 cells.

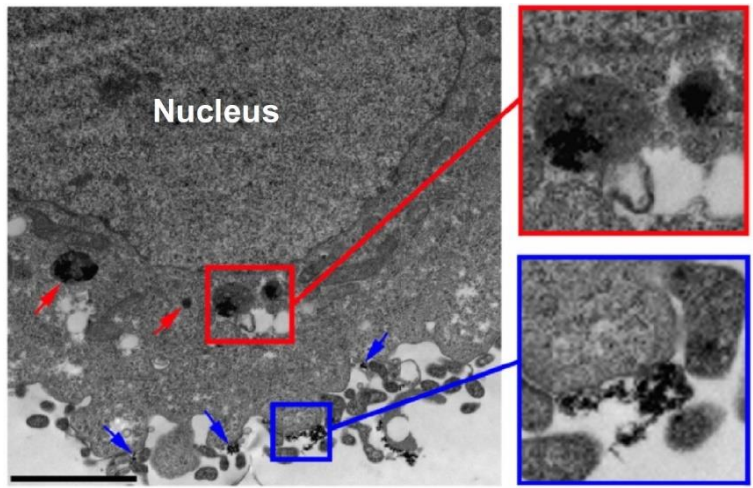

**Figure 13.** Transmission electron microscopy of the MCF-7 cells incubated with SPIONs.

**Notes:** Following incubation with SPIONs (150 µg/mL) for 24 hours, nanoparticles could be detected as being attached to the cell membrane (blue arrows) and incorporated into the endosome-like structures in the cytoplasm (red arrows). Scale bar: 2 µm.

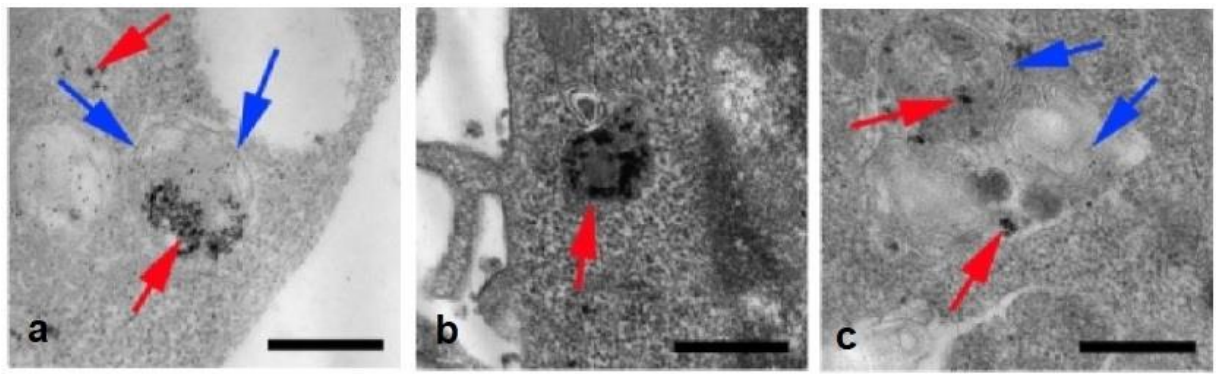

**Figure 14.** TEM of MCF-7 cells incubated for 24 hours with SPION-DOX conjugates.

**Notes:** (a) TEM of MCF-7 cells incubated for 24 hours with bare iron-oxide nanoparticles. Electron-dense nanoparticles were present in the cytoplasm of cells in endosome-like structures (red solid arrow). Scale bar: 500 nm. (b), (c) TEM of cells incubated with SPIONs-DOX conjugates for 24 hours.

**In vitro toxicity of doxorubicin loading onto the SPIONs**

$IC_{50}$ and $IC_{90}$ analyses of MCF-7 and MDA-MB-231 cell lines, after 24, 48, 72, 96 h exposure to DOX-loaded polycarboxylic-coated SPIONs were recorded at 1.9 ± 0.5 and 4.2 ± 0.7 mM concentration of drug, while, $IC_{90}$ of both cell lines was seen at 4.3 ± 0.7 and 6.8 ± 1.0 mM DOX concentration after 96 h for MCF-7 and MDA-MB-231 cells respectively. Inhibition concentrations for DOX-loaded Rhodamine B-SPIONs were recorded at 2.0 ± 0.6 and 3.9 ± 0.8 mM concentration of drug for $IC_{50}$, while, $IC_{90}$ of both cell lines was seen at 4.0 ± 0.8 and 6.0 ± 1.2 mM DOX concentration after 96 h for MCF-7 and MDA-MB-231 cells respectively. This demonstrates that the MDA-MB-231 cells were more resistant to the lower concentrations of drug as compared to the MCF-7 cells (Table 3; $p < 0.05$).

**Table 3:** $IC_{50}$ and $IC_{90}$ of MCF-7 and MDA-MB-231 cell lines after 24, 48, 72, 96 exposure to DOX-loaded n(COOH)-SPIONs

| | 24 h | 48 h | 72 h | 96 h |
|---|---|---|---|---|
| **$IC_{50}$ for DOX-loaded n(COOH)-SPIONs (mM)** | | | | |
| Type of cell line | | | | |
| MCF-7 | 4.0 ± 0.9 | 3.1 ± 0.8 | 2.4 ± 0.7 | 1.9 ± 0.5 |
| MDA-MB-231 | 6.4 ± 1.8 | 5.3 ± 1.2 | 4.8 ± 0.9 | 4.2 ± 0.7 |
| **$IC_{90}$ for DOX-loaded n(COOH)-SPIONs (mM)** | | | | |
| Type of cell line | | | | |
| MCF-7 | 8.9 ± 1.5 | 7.6 ± 1.0 | 6.1 ± 0.6 | 4.3 ± 0.7 |
| MDA-MB-231 | 10.6 ± 3.4 | 9.4 ± 2.1 | 7.3 ± 0.8 | 6.8 ± 1.0 |

**Table 4:** $IC_{50}$ and $IC_{90}$ of MCF-7 and MDA-MB-231 cell lines after 24, 48, 72, 96 exposure to DOX-loaded Rhodamine B-SPIONs

| | 24 h | 48 h | 72 h | 96 h |
|---|---|---|---|---|
| **$IC_{50}$ for DOX-loaded Rhodamine B-SPIONs (mM)** | | | | |
| Type of cell line | | | | |
| MCF-7 | 3.8 ± 0.9 | 3.0 ± 0.7 | 2.3 ± 0.6 | 2.0 ± 0.6 |
| MDA-MB-231 | 6.7 ± 1.4 | 5.6 ± 1.0 | 4.6 ± 0.8 | 3.9 ± 0.8 |
| **$IC_{90}$ for DOX-loaded Rhodamine B-SPIONs (mM)** | | | | |
| Type of cell line | | | | |
| MCF-7 | 8.7 ± 1.3 | 7.4 ± 1.1 | 5.9 ± 0.7 | 4.0 ± 0.8 |
| MDA-MB-231 | 10.9 ± 2.7 | 9.6 ± 2.0 | 7.7 ± 1.3 | 6.0 ± 1.2 |

DOX, doxorubicin. The $IC_{50}$ and $IC_{90}$ are defined as the concentrations causing 50% and 90% growth inhibition in treated cells, respectively, when compared to control cells. Values are means ± SEM of at least three separate experiments.

**DOX release profile**

The DOX release profiles of n(COOH)-SPIONs and Rhodamine B-SPIONs were analyzed to check the free and the conjugated drug, at the pH range of 1.5-7.0. At Ph 7.0, a small amount of the drug release was observed after the incubation period of 48 h.

The DOX release profiles of the free and the conjugated drug, at the pH range of 1.5-7.0. At pH 7.0, a small amount of the drug release was observed after the incubation period of 48 h. This is a desirable characteristic as the pH 7.4 is the undesired pH for the proper release of the drugs from the nanoconjugated drug carrier. This will also prevent the premature release of the drugs before the nanoconjugates reach the cancer cells. It shows that the pH 6.0 provided the desirable conditions for the proper drug release. The first 10 h represent the period of initial rapid release, followed by a steady state.

This pH-dependent drug release behavior is favorable for the chemotherapeutic process as it can significantly reduce the preterm drug release on the body pH level (pH 7.4) and maximizing the amount of drug reaching the target tumor cells, once the drug-loaded SPIONs internalize and enter the tumor by endocytosis (pH 4.5-6.5) (Figure 15).

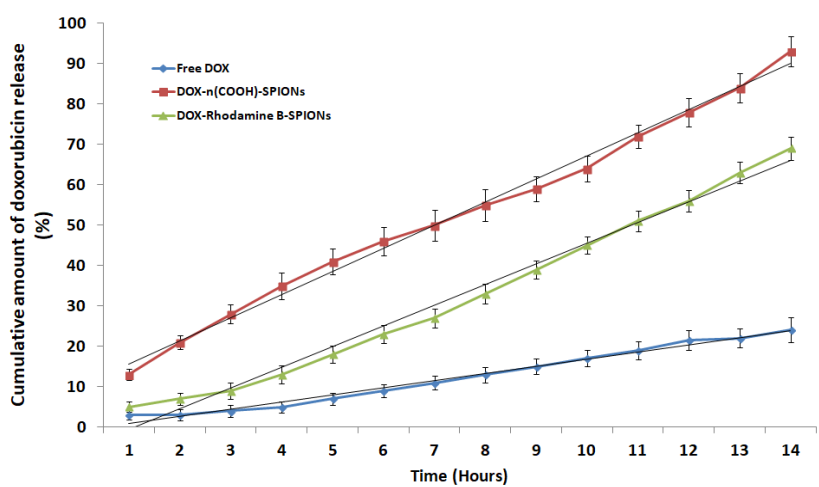

**Figure 15.** DOX release profiles of the free and the conjugated drug.

**Discussion**

The aim of the study was to couple a second generation of superparamagnetic nanoparticles with doxorubicin, a very effective chemotherapeutic agent used for breast cancer therapy. Due to the well-known toxicity of Doxorubicin, new ways to deliver the drug into target tissues are urgently needed.

The magnetic core of the nanoparticles was designed to induce controlled magnetic hyperthermia by an external magnetic field. Superparamagnetic iron oxide MNPs were successfully synthesized by co-precipitation method that is facile and versatile ease of synthesizing. The great advantage of this synthesis method is high yield in the production of nanoparticles in a short time.

The pursuit of innovative, multifunctional, more efficient, and safer treatments is a major challenge in preclinical nanoparticle-mediated thermotherapeutic research for breast cancer. Here, we reported that iron oxide nanoparticles have the dual capacity to act as both magnetic and drug delivery agents.

Doxorubicin-loaded polycarboxylic acid-coated superparamagnetic iron oxide nano-particles ($Fe_3O_4$; n(COOH)-SPIONs) with an average particle size of 30 nm, with high encapsulation efficiencies, were obtained by the electrostatic loading of doxorubicin (DOX) to SPIONs through the electrostatic interaction between the carboxyl group of carboxylic acid functionalized iron-oxide nanoparticles and the amine group of DOX. Iron oxide-based magnetic nanoassemblies are a family of drug delivery systems with high potential like theranostic due to their low toxicity, size-dependent superparamag-netism, large surface area, and biocompatibility. In this study, we tested a novel DOX nanocarrier and its effects on normal and breast cancer cell lines.

In order to verify the targeting efficacy of nanoparticles, cell viability evaluation was conducted to estimate the viability of MCF-7 and MDA-MB-231 cells related to the in vitro efficacy of DOX, DOX-SPIONs, and DOX-SPIONs MHT with DOX concentration of 5 µM for 24, 48 and 72 hours. A gradually time-dependent decreasing trend in the cell viability with all treatments was observed. We found that the cell viability of DOX-SPIONs-treated cells was lower than that of only DOX-treated cells, and the difference was statistically significant; this may be explained by the action of DOX-SPIONs con-jugates which can be easily translocated across the plasma membrane by endocytosis into breast cancer cells to exert their inhibitory effect on the cells. The viability of the DOX-SPIONs-treated cells was lower than that of the other two groups ($p<0.05$). The active targeting of conjugated nanoparticles led to more successful transport through the cellular membrane. The half inhibitory concentration of DOX-SPIONs at 24 hours was ~4.0 and ~6.4 mM for MCF-7 and MDA-MB-231 cells, respectively. Additionally, DOX released from the DOX-SPIONs was measured through dialysis at range of pH between 1.5 and 7.0 (Figure 3). Under conditions including a neutral dialysate (pH =7.4), DOX-SPIONs displayed a slow and sustained DOX release rate of ~50% within 7 hours, with ~90% of DOX released within 14 hours. These results show that DOX-SPIONs ensure a smooth and continuous diffusion of DOX in a neutral environment.

The comparison of the results of this study according to literature showed a promising ability of doxorubicin-conjugated iron-oxide nanoparticles to promote targeted apoptosis of cancer cells and a potential to act as an antimetastatic chemothermotherapeutic agent [11].

This work aimed to implement an innovative multi-therapeutic strategy that combines the use of superparamagnetic iron oxide nanoparticles properties for dual tumor targeting therapy on cancer cells by exploiting magnetic induced hyperthermia and selective drug delivery to target cancer. One factor that can contribute to nanotoxicity is the size of the particles. Smaller particles have a greater reactive surface area, are more chemically reactive, and produce higher numbers of ROS than larger particles. This is one of the primary mechanisms of nanotoxicity and it may result in the oxidative stress causing inflammation and damage to proteins, membranes, and DNA. The size and surface charge of NPs are important physiochemical parameters in designing drug delivery vehicles for therapeutic applications. Transition metal ions such as Fe(II) and Fe(III) can generate ROS. Superoxide, produced via normal metabolic processes and redox cycling within the cell, reacts with hydrogen peroxide to form hydroxyl free radicals. The catalysts in the present case are the iron ions on the surface of SPIONs. Free radicals generated by SPIONs or DOX-SPIONs are responsible for oxidative stress and cell damage (lipid peroxidation and protein oxidation).

The major mechanism of action of free DOX involves its intercalation in the DNA and inhibition of topoisomerase II. Several secondary mechanisms of DOX action are described as DOX-induced production of ROS. Mitochondrially located ROS can also be detected rapidly after DOX administration, most probably due to respiratory chain-produced superoxide radicals. One feature of DOX is to produce ROS via one-electron reduction to the corresponding semiquinone free radicals that then react rapidly with oxygen to generate superoxide radical anions.

In our research, DOX-SPIONs showed greater targeted cytotoxicity than free DOX on breast cancer cells, regardless of DOX concentration. Viability was decreased with increasing DOX concentration. High cytotoxicity is based on successful delivery of DOX into the nuclei of cancer cells. The treatment with DOX alone produced a very substantial decrease in cell viability. The incorporation of DOX into SPIONs core strongly enhanced the cytotoxic effect of drug when compared with the free drug.

The aim of this study was to investigate the application of SPIONs in magnetic hyperthermia of MCF7 cancer cells.

The results demonstrated the potential of the DOX-n(COOH)-SPIONs and RhB-SPIONs to achieve dual tumor targeting and magnetic-induced drug delivery by magnetic field-guided in breast cancer cells. This may reduce the required dose of doxorubicin and consequently reduction the side effects of this drug. We discovered an intrinsic therapeutic effect of magnetic nanoparticles on cancer growth of breast cancer cells. We produced a novel innovative nanoparticle with multifunctional feature for magnetic targeted tumor diagnosis and anticancer therapy. The nanoplatform produced exhibits the high magnetization, excellent stability as well as high efficiency of targeted drug delivery. Tumor targeting can be obtained under the guidance of an external magnetic field, which improves the specific delivery of nanoparticles to the tumor sites and thus improve the therapeutic efficacy greatly. So, we have demonstrated that DOX-SPIONs can promote apoptosis and lessen the negative effects of chemotherapeutic agents on control cells mammary epithelial cells during the in vitro conditions because they have a targeted cytotoxic effect on cancer cell lines. To optimize this drug delivery system, greater understanding of the different mechanisms of biological interactions, and particle engineering, is still required. Further advances are needed in order to turn the concept of nanoparticle technology into a realistic practical application as the next generation of drug delivery system.

**Conclusion**

In this study, polycarboxylic and Rhodamine-B iron-oxide nanoparticles were successfully synthesized. Our current study indicated the clear advantages of DOX-SPIONs in comparison with free DOX or bare n(COOH)-SPIONs for the treatment of breast carcinoma without discernible side effects. Furthermore, the established and relevant MRI technology could be readily used to orientate DOX-SPIONs for preferable diagnosis and treatment of breast cancer. DOX-SPIONs equipped with dual targeting methods and targeted drug delivery capacity are a promising theranostic candidate for breast cancer. According to these results, the prepared nanoparticles are efficient and appropriate theranostics tools for breast cancer cells.

In conclusion, this study indicates that SPIONs are promising therapeutic agents for magnetic hyperthermia of breast cancer cells and offers a dual magnetotherapeutic approach with iron oxide nanoparticles as the sole heat mediators. We demonstrated that iron oxide nanoparticles can be remotely activated with an alternating magnetic field, achieving a very efficient heat conversion. Remarkably, the dual magnetic and drug delivery action resulted in complete cell death in vitro at low iron doses, tolerable magnetic field and frequency conditions. This cumulative, if not synergistic, heat therapy is thus promising for tumor treatment with minimal collateral tissue damage.

DOX-SPIONs for targeted cancer chemotherapy are exocytosed from tumor cells after incubation. In summary, we have successfully developed biocompatible magnetic nanoparticles platform for targeted cancer chemotherapy. Our study clearly demonstrates that iron-oxide nanoparticles have potential as drug carriers to improve the anticancer efficacy.

**Future directions**

The therapeutic approaches based on enhanced permeability and retention (EPR) effect can sustain an increased nanoparticles-based drug delivery system for cancer therapy. The external field triggered nanotheranostics (therapy + diagnostics) approach together with can allow precise control of the drug release, timing, intensity, duration of drug delivery and treatment monitoring. This means a boost to the treatment all-in-one efficacy.

This can be observed like a progression for convergence therapy with real-time monitoring of tumor destruction. When tumor cells are incubated with doxorubicin-loaded magnetic nanoparticles the tumor cells can retain for longer time the nanoparticles. DOX-SPIONs could exhibit enhanced tumor accumulation, tumor penetration and cross-reactive cellular uptake, resulting in augmented DOX enrichment in total tumor cells and side population cells.

Several physical energy sources could be used in future to trigger hybrid nanostructures. In recent times various research groups have invented novel ways to utilize nanoscale materials that are capable of responding to external physical stimulus such as (i) light (ii) magnetic fields (iii) ultrasound (iv) radiofrequencies (v) x-rays. Novel nanostructures are highly sensitive to external stimulus and can perform 'on demand' cancer theranostics.

The perspective use of iron-oxide nanoparticles in nanotheranostics application can be a perspective use and application for targeted use of medical physics solutions to contrast cancer advancement.

**Acknowledgements**

The research leading to these results has received funding from the European Union Seventh Framework Programme (FP7-PEOPLE-2013-COFUND) under grant agreement n° 609020 - Scientia Fellows.

This article is based upon work from COST Action CA 17140 "Cancer Nanomedicine from the Bench to the Bedside" supported by COST (European Cooperation in Science and Technology).

**References**

1. Malvezzi M, Carioli G, Bertuccio P, Boffetta P, Levi F, La Vecchia C, Negri E. European cancer mortality predictions for the year 2019 with focus on breast cancer. Ann Oncol. 2019 May 1;30(5):781-787. doi: 10.1093/annonc/mdz051.

2. Morris SA, Farrell D, Grodzinski P. Nanotechnologies in cancer treatment and diagnosis. J Natl Compr Canc Netw. 2014; 12(12):1727-33.

3. Kumar L, Harish P, Malik PS, Khurana S. Chemotherapy and targeted therapy in the management of cervical cancer. Curr Probl Cancer. 2018; pii: S0147-0272 (18) 30019-9.
4. Lüpertz R, Wätjen W, Kahl R, Chovolou Y. Dose- and time-dependent effects of doxorubicin on cytotoxicity, cell cycle and apoptotic cell death in human colon cancer cells. Toxicology. 2010; 271:115-121.
5. Yeh ET, Chang HM. Oncocardiology-Past, Present, and Future: A Review. JAMA Cardiol. 2016 Dec 1;1(9):1066-1072.
6. Menna P, Paz OG, Chello M, Covino E, Salvatorelli E, Minotti G. Anthracycline cardiotoxicity. Expert Opin Drug Saf. 2012; 11 Suppl 1: S21-36. doi: 10.1517/14740338.2011.589834.
7. Vejpongsa P, Yeh ET. Prevention of anthracycline-induced cardiotoxicity: challenges and opportunities. J Am Coll Cardiol. 2014; 64(9):938-45. doi: 10.1016/j.jacc.2014.06.1167.
8. Gautier J, Munnier E, Paillard A, et al. A pharmaceutical study of doxorubicin-loaded PEGylated nanoparticles for magnetic drug targeting. Int J Pharm. 2012; 424:16-25.
9. Nedyalkova M, Donkova B, Romanova J, Tzvetkov G, Madurga S, Simeonov V. Iron oxide nanoparticles - In vivo/in vitro biomedical applications and in silico studies. Adv Colloid Interface Sci. 2017; 249:192-212.
10. Wang Y, Sun S, Zhang Z, Shi D. Nanomaterials for Cancer Precision Medicine. Adv Mater. 2018.
11. Espinosa A, Di Corato R, et al. Duality of Iron Oxide Nanoparticles in Cancer Therapy: Amplification of Heating Efficiency by Magnetic Hyperthermia and Photothermal Bimodal Treatment. ACS Nano. 2016; 10(2):2436-46.
12. He YP, Miao YM, Li CR, Wang SQ, et al. Phys. Rev. B, 2005, 71(1-9), 125411.
13. Kim Y II, Kim D, and Lee CS. Phys. B, 2003, 337, 42-51.
14. F. Grasset, N. Labhsetwar, D. Li, D. C. Park, N. Saito, H. Haneda, O. Cador, T. Roisnel, S. Mornet, E. Duguet, J. Portier and J. Etourneau, Langmuir, 2002, 18, 8209-8216.
15. A. Schlossbauer, D. Schaffert, J. Kecht, E. Wagner and T. Bein, J. Am. Chem. Soc., 2008, 130, 12558-12559.
16. L. Josephson, M. F. Kircher, U. Mahmood, Y. Tang and R. Weissleder, Bioconjugate Chem., 2002, 13, 554-560.
17. Maier-Hauff K, Ulrich F, Nestler D, et al. Efficacy and safety of intratumoral thermotherapy using magnetic iron-oxide nanoparticles combined with external beam radiotherapy on patients with recurrent glioblastoma multiforme. J Neurooncol. 2011;103(2):317-324.
18. Chen Q, Ke H, Dai Z and Liu Z. Nanoscale theranostics for physical stimulus responsive cancer therapies Biomaterials 2015; 73, 214-30.
19. Sneider A, Van Dyke D, Paliwal S and Rai P 2017 Remotely triggered nanotheranostics for cancer applications. Nanotheranostics 2017; 1, 1-22.